\documentclass[prc,preprint,preprintnumbers,superscriptaddress,nofootinbib]{revtex4}

\usepackage{amsmath,slashed}
\usepackage{graphicx,graphics,xcolor}
\usepackage{dcolumn}
\usepackage[hyperfootnotes=false]{hyperref}
\usepackage{xspace}
\usepackage{enumerate}
\usepackage{varwidth}
\usepackage{comment}
\usepackage{physics}

\usepackage{amssymb}
\usepackage{mathtools}
\usepackage{dsfont}
\usepackage{multirow}
\usepackage{float}

\newcommand{\beq}{\begin{equation}}
\newcommand{\eeq}{\end{equation}}

\newcommand{\inner}[2]{\langle #1 | #2 \rangle}
\newcommand{\avg}[1]{\langle #1 \rangle}

\allowdisplaybreaks[1]

\begin{document}

\preprint{LA-UR-26-23797}

\title{Quantum Monte Carlo calculation of $\delta_C$ in the superallowed beta decay of $^{10}$C}

\author{Maria Piarulli}
\affiliation{Department of Physics, Washington University in Saint Louis, Saint Louis, MO 63130, USA}
\affiliation{McDonnell Center for the Space Sciences at Washington University in St. Louis, MO 63130, USA}

\author{R. B. Wiringa}
\affiliation{Physics Division, Argonne National Laboratory, Argonne, IL 60439, USA}

\author{Alessandro Lovato}
\affiliation{Physics Division, Argonne National Laboratory, Argonne, IL 60439, USA}
\affiliation{Computational Science Division, Argonne National Laboratory, Argonne, Illinois 60439, USA}
\affiliation{INFN-TIFPA Trento Institute for Fundamental Physics and Applications, Trento, Italy}
\affiliation{Instituto de Física Corpuscular (IFIC), Consejo Superior de Investigaciones Científicas (CSIC) and Universidad de Valencia
E-46980 Paterna, Valencia, Spain}

\author{Garrett B. King}
\affiliation{Theoretical Division, Los Alamos National Laboratory, Los Alamos, NM 87545, USA}

\author{Saori Pastore}
\affiliation{Department of Physics, Washington University in Saint Louis, Saint Louis, MO 63130, USA}
\affiliation{McDonnell Center for the Space Sciences at Washington University in St. Louis, MO 63130, USA}

\begin{abstract}
We perform an \textit{ab initio} quantum Monte Carlo calculation of the isospin-symmetry-breaking correction $\delta_C$ to the superallowed $\beta$ decay of $^{10}{\rm C}$. Using both phenomenological and chiral nuclear interactions, we evaluate the Fermi matrix element and quantify its deviation from the canonical $\sqrt{2}$ value. The resulting $\delta_C$ values lie in the range $\approx 0.15$--$0.25\%$ and are consistent, within sizable uncertainties (approximately $34\%$--$65\%$ relative), across Hamiltonians, indicating no statistically significant dependence on the choice of nuclear interaction. The extracted values of $V_{ud}$ are also found to be compatible with current determinations within these uncertainties.
\end{abstract}

\maketitle

\tableofcontents

\section{Introduction}
Superallowed $0^+ \!\to 0^+$ Fermi $\beta$ decays provide one of the most stringent low-energy tests of the Standard Model. After applying radiative and nuclear-structure--dependent corrections, the corrected comparative half-lives $\mathcal{F}t$ are expected to be nucleus independent \cite{Hardy:2014qxa,Hardy:2020qwl}, enabling a precise determination of the vector coupling constant, $G_V$, and the $V_{ud}$ element of the Cabibbo--Kobayashi--Maskawa (CKM) matrix \cite{Cabibbo:1963yz,Kobayashi:1973fv}. Together with kaon and neutron decay data, these results play a central role in tests of CKM unitarity and in searches for physics beyond the Standard Model \cite{Towner:2007np,Marciano:2005ec,Seng:2018yzq,Seng:2018qru}. At the current level of precision, such studies probe the Standard Model at the level of a few parts in $10^{-4}$, making them sensitive to even small theoretical uncertainties.

The conventional master formula for the corrected comparative half-life $\mathcal{F}t$ value between $T=1$ analog states is given by~\cite{Hardy:2020qwl},
\begin{equation}
    \mathcal{F}t = ft(1+\delta_{\rm NS} - \delta_C)(1+\delta'_R) = \frac{K}{G_F^2 |V_{ud}|^2 |M^0_F|^2 (1 + \Delta^V_R) }\, , \label{eq:master}
\end{equation} 
where $ft$ is the experimentally determined comparative half-life, $K=2\pi^3\hbar^7\ln(2)/(m_e^5c^4)$, $G_F$ is the Fermi coupling constant, and $|M^0_F|^2=2$ is the Fermi matrix element between states in the isotriplet. $\Delta^V_R$ is a transition-independent radiative correction, while $\delta'_R$ and $\delta_{\rm NS}$ are nucleus-dependent radiative corrections. The former is sensitive to the global properties of the nucleus, while $\delta_{\rm NS}$ depends on the particular structure of the nucleus  and encodes short- and intermediate-range nuclear effects in electroweak radiative processes~\cite{Towner:2007np,Marciano:2005ec}. In addition to these radiative corrections, the correction, $\delta_C$, arises from isospin mixing in the nuclear wave functions primarily induced by the Coulomb interaction and other charge-dependent components of the nuclear force. Recently, the authors of Ref.~\cite{Plestid:2025idc} revisited this correction from the perspective of an LSZ-based perturbative framework for QED effects in composite systems, showing how the conventional intranuclear-Coulomb contribution to $\delta_C$ emerges as part of the two-loop structure of the theory.

At the current level of precision, the uncertainty in $V_{ud}$ is generally dominated not by experiment but by theory, in particular by nuclear-structure--dependent correction ($\delta_{\rm NS}-\delta_C$) applied to the measured decay rates \cite{Hardy:2020qwl}. Both nuclear structure corrections are essential for achieving the sub-percent theoretical precision required for meaningful CKM unitarity tests. A great deal of effort has recently been focused on $\delta_{\rm NS}$, particularly, after the authors of Ref.~\cite{Seng:2018qru} suggested that the traditional evaluation of this quantity was incomplete. An alternative formalism was proposed in Refs.~\cite{Cirigliano:2024msg,Cirigliano:2024rfk} based on effective field theory, and {\it ab initio} calculations of $\delta_{\rm NS}$ were performed for $^{10}$C~\cite{Gennari:2024sbn,King:2025fph} and $^{14}$O~\cite{Cirigliano:2024msg}. The status of these calculation was recently reviewed in Ref.~\cite{Sargsyan:2026ygi}. 

For the determination of $\delta_C$, there have been proposals to extract this quantity from electroweak radii and isospin symmetry breaking observables~\cite{Seng:2022epj,Seng:2023cvt}, as well as recent refinements to shell-model approaches to compute this quantity~\cite{Xayavong:2025jdh}. An {\it ab initio} calculation of $\delta_C$ would clearly complement these efforts. Further, such a calculation would supplement recent studies of $\delta_{\rm NS}$ by providing both corrections in a consistent framework. In recent work, we performed an {\it ab initio} quantum Monte Carlo (QMC) calculation of $\delta_{\rm NS}$ in $^{10}$C \cite{King:2025fph}. That study demonstrated that QMC methods, combined with realistic two- and three-nucleon interactions, can provide controlled and systematically improvable evaluations of electroweak nuclear corrections beyond traditional shell-model approaches.

In this work, we extend this {\it ab initio} program by presenting a QMC calculation of the isospin-symmetry--breaking correction $\delta_C$ for the superallowed $\beta$ decay of $^{10}$C. The $^{10}$C $\to$ $^{10}$B* transition is among the lightest superallowed emitters and has long been recognized as a sensitive benchmark for studies of isospin breaking in nuclei \cite{Hardy:2014qxa,Towner:2007np}. Its relatively simple structure, combined with high-precision experimental data, makes it an ideal system for assessing theoretical treatments of $\delta_C$. 

The current evaluation of $\delta_C$ relies primarily on shell-model or mean-field--based calculations, which necessarily involve model assumptions and phenomenological adjustments \cite{Towner:2007np,Hardy:2020qwl}. In contrast, QMC methods explicitly include correlations induced by realistic nuclear interactions and provide a natural framework for studying isospin breaking in light nuclei from first principles \cite{Pudliner:1997ck,Pieper:2001mp,Wiringa:2013fia,Piarulli:2017dwd}.
In particular, we perform Green's function Monte Carlo (GFMC) calculations of $\delta_C$. The GFMC framework introduces many-body correlations by removing excited state contamination in a trial wave function via the exponential form of the solution to the imaginary-time Schr\"{o}dinger Equation. In the present calculation, the imaginary-time evolution is performed in the charge basis, rather than in the good-isospin basis, and includes both charge-dependent (CD) and charge-symmetry-breaking (CSB) terms in the Hamiltonian. The most important of these terms arises from the Coulomb repulsion between finite-size protons. This approach breaks isospin symmetry during propagation and allows for a controlled extraction of $\delta_C$.

Together with the recent GFMC determination of $\delta_{\rm NS}$ in Ref.~\cite{King:2025akz}, the present work represents an important step toward a fully {\it ab initio} description of all nuclear-structure--dependent corrections entering superallowed $\beta$ decay. Such calculations are essential for understanding theoretical uncertainties in $V_{ud}$, informing the current status of CKM unitarity, and strengthening the role of nuclear theory in precision tests of the Standard Model.

This paper is organized as follows. In Sec.~\ref{sec:hamiltonian}, we describe the nuclear Hamiltonians employed in this work, including both phenomenological and chiral two- and three-nucleon interactions. In Sec.~\ref{sec:qmc}, we outline the QMC methods used to compute the nuclear wave functions and transition matrix elements, including both variational Monte Carlo (VMC) and Green’s function Monte Carlo (GFMC) approaches. The results for energies, radii, isospin admixtures, and the Fermi matrix element are presented in Sec.~\ref{sec:results}, where we also discuss the extraction of $\delta_C$ and the corresponding implications for $V_{ud}$, including their dependence on the nuclear interaction. Finally, conclusions and perspectives are given in Sec.~\ref{sec:conclusions}.

\section{Nuclear Hamiltonians}\label{sec:hamiltonian}

The nuclear Hamiltonians used in this work have the general form,
\begin{equation}
 H = \sum_{i} T_i + {\sum_{i<j}} v_{ij} + \sum_{i<j<k}
V_{ijk} \ ,
\end{equation}
where $T_i$ is the single-nucleon nonrelativistic kinetic energy and $v_{ij}$ and $V_{ijk}$ are two- and three-nucleon potentials, respectively.
The nuclear interaction is predominantly charge-independent (CI) but here we are interested in small sources of charge-independence breaking (CIB), both charge-dependent (CD) and charge-symmetry breaking (CSB) terms. 
The kinetic energy includes both CI and CSB contributions, the latter coming from the neutron-proton mass difference:
\begin{eqnarray}
  T_{i} & = & T^{\rm CI}_{i} + T^{\rm CSB}_{i} \\
        & = & -\frac{\hbar^2}{4}
     (\frac{1}{m_{p}} + \frac{1}{m_{n}}) \nabla^{2}_{i}
    -\frac{\hbar^2}{4} (\frac{1}{m_{p}} - \frac{1}{m_{n}})\tau_{iz}
     \nabla^{2}_{i} \nonumber \ ,
\end{eqnarray}
where $\tau_{iz}$ is the third component of isospin for nucleon $i$, and $m_p$ and $m_n$ are the proton and neutron masses, respectively.
We use two different potential models which include explicit charge-independence breaking: the phenomenological Argonne $v_{18}$ (AV18)~\cite{Wiringa:1994wb} two-body interaction with either the Urbana X (UX)~\cite{Wiringa:2014} or Illinois-7 (IL7)~\cite{Carlson:2014vla} three-body interactions -- referred to collectively as the AV18+UX and AV18+IL7 models -- and two versions of the chiral effective field theory (EFT) Norfolk two- and three-nucleon interaction (NV2+3)~\cite{Piarulli:2014bda,Piarulli:2016vel,Piarulli:2017dwd,Baroni:2018fdn}. 

The AV18 potential includes a full long-range electromagnetic (EM) interaction, the dominant one-pion exchange (OPE) contributions, intermediate-range contributions that approximate two-pion exchange (TPE), and a short-range potential whose shape is given by a modified Woods-Saxon form~\cite{Wiringa:1994wb}; 
\begin{equation}
  v_{ij} = v_{\gamma}(r_{ij})
         + \sum_{p=1,18} [ v^p_{\pi}(r_{ij}) + v^p_{I}(r_{ij})
         + v^p_{S}(r_{ij}) ] O^p_{ij} \ .
\end{equation}
The electromagnetic interaction $v_\gamma$ includes Coulomb, magnetic moment, and vacuum polarization terms; it is expressed in terms of $pp$, $np$, and $nn$ pair projection operators.
The nuclear portion has 18 different operator structures, including 14 CI terms:
\begin{eqnarray}
  O^{p=1,14}_{ij} & = & [1, \sigma_{i}\cdot\sigma_{j}, S_{ij},
    {\bf L\cdot S},{\bf L}^{2},{\bf L}^{2}(\sigma_{i}\cdot\sigma_{j}),
    ({\bf L\cdot S})^{2}] 
                   \otimes [1,\tau_{i}\cdot\tau_{j}] \ ,
\label{eq:ops}
\end{eqnarray}
plus three CD terms and one CSB term:
\begin{equation}
O^{p=15,18}_{ij} = [1, \sigma_{i}\cdot\sigma_{j},
S_{ij}]\otimes T_{ij} \ , \ (\tau_i+\tau_j)_z \ .
\end{equation}
Here $\sigma_{i}$ is the Pauli spin operator for nucleon $i$, $S_{ij} = 3(\sigma_{i}\cdot {\hat r}_{ij})(\sigma_{j}\cdot {\hat r}_{ij}) - \sigma_{i}\cdot\sigma_{j}$ is the tensor operator, ${\bf S} = (\sigma_{i} + \sigma_{j})/2$ is the total pair spin, ${\bf L}$ is the pair orbital momentum operator, and $T_{ij} = 3\tau_{iz}\tau_{jz} - \tau_{i}\cdot\tau_{j}$ is the isotensor operator.
The long-range OPE has a significant CD contribution arising from the mass difference between charged and neutral pions.
The intermediate and short-range contributions to the force are constrained by the differences between the considerable amount of $pp$ and $np$ scattering data in the $^1S_0$ channel.
There are 42 parameters in the potential constrained with 4301 $np$ and $pp$ scattering data from the Nijmegen partial wave analysis~\cite{Stoks:1993tb} with a $\chi^2$/datum of ${\sim}1.1$, as well as the deuteron binding energy and $nn$ scattering length.
The strength of the CSB term is determined by the difference between $pp$ and $nn$ scattering lengths.
Extended versions of the CIB terms in AV18 have been investigated but are not used here~\cite{Wiringa:2013fia}.

The UX is a CI phenomenological model of the three-nucleon force that hybridizes two other models; namely, the IL7 and Urbana IX (UIX) models~\cite{Carlson:2014vla}. The UX supplements the long-range two-pion P-wave and central short-range repulsion of the UIX with a two-pion S-wave term, taking the strengths of all three of these terms from the IL7 parametrization. The IL7 includes three-pion ring diagrams involving one or two intermediate $\Delta$'s, in addition to the terms of the UX model, and is fit to ground state and low-lying excitation energies of $A\le 10$ nuclei.  The three-nucleon potential can also have CD contributions if the charged and neutral pion mass differences are kept in the two-pion exchange terms, but they have been neglected here.

The NV2+3 interaction is a semi-phenomenological model derived within a chiral EFT framework that retains nucleons, pions, and
$\Delta$-isobars as explicit degrees of freedom. This approach exploits the
natural separation between the typical momentum scale of nucleons in nuclei,
denoted by a generic low-energy scale $Q \sim M_\pi$ (or the characteristic
nucleon momentum in nuclei), as well as the $N$--$\Delta$ mass splitting, and
the larger chiral symmetry--breaking scale $\Lambda_\chi$ associated with
heavier degrees of freedom. As a result, the latter can be integrated out,
leading to an effective description in terms of an expansion in powers of the
chiral parameter $\varepsilon_\chi \equiv Q/\Lambda_\chi$. While such an
expansion is, in principle, systematic, the detailed implementation of power
counting remains an open issue~\cite{Hammer:2019poc}, and a specific
organizational scheme must therefore be adopted.
The NV2 potential was derived by first developing a minimally nonlocal two-body force by means of Fierz transformations up to N$^3$LO ({\it i.e.}, to $\mathcal{O}(\varepsilon_\chi^4)$)~\cite{Piarulli:2014bda} in the power counting prescription of Weinberg~\cite{Weinberg:1991um}. Because the use of local potentials are more efficacious for QMC approaches~\cite{Carlson:2014vla}, only those local terms at N$^3$LO necessary to provide a good description of nucleon-nucleon scattering were retained to form the semi-phenomenological NV2 model~\cite{Piarulli:2016vel}.
This is supplemented by an N$^2$LO ($\mathcal{O}((\varepsilon_\chi^3)$) three-body force, the NV3, based on terms first derived by van Kolck {\it et al.}~\cite{vanKolck:1994yi} and Epelbaum {\it et al.}~\cite{Epelbaum:2002vt}. In the $\Delta$-full picture, the diagram involving the excitation of an intermediate $\Delta$-- corresponding to the Fujita-Miyazawa (FM) term~\cite{Fujita:1957zz}-- is promoted to NLO ($\mathcal{O}(\varepsilon_\chi^2)$) in the chiral expansion~\cite{Piarulli:2019cqu}. In total, the nucleon-nucleon NV2 interactions have 26 unknown low-energy constants (LECs) that parameterize their short-range behavior. In addition, there are two additional unknown LECs in the three-body sector. Various fitting schemes were used to obtain a suite of NV2+3 interactions; namely, choices for which two-nucleon data to fit, how to regularize singularities in the long-range terms of the potential, and how to fit the three-body force. In this work, we focus on two interactions -- the NV2+3-Ia and NV2+3-Ia$^{\star}$ -- which fit the NV2-Ia two-body force with the same data and regularization scheme, but differ in how the three-body force was fit. In particular, the two-body force is fit to ${\sim}$2700  $np$ and $pp$ scattering data from the Granada database~\cite{Perez:2013jpa,Perez:2013oba,Perez:2014yla}, and like the AV18, has a $\chi^2/$datum of ${\sim}1.1$. The NV3 fits the three-body force to the trinucleon binding energies and the $n-d$ doublet scattering length~\cite{Piarulli:2017dwd}, while the NV3$^{\star}$ was fit to the trinucleon binding energies and the tritium Gamow-Teller $\beta$ decay matrix element~\cite{Baroni:2018fdn}. The NV2-Ia combined with the former fitting scheme is denoted as the NV2+3-Ia interaction, while the combination with the latter fitting scheme is called NV2+3-Ia$^{\star}$~\cite{Piarulli:2016vel}.

The NV2 models have essentially the same long-range electromagnetic potential $v_\gamma$ as the AV18, and the same operator structure for the nuclear part with two exceptions: there are no ${\bf L}^{2}(\sigma_{i}\cdot\sigma_{j})$ terms in the CI part, while one CD term ${\bf L\cdot S} \otimes T_{ij}$ has been added.
This latter term is constrained by differences in the $pp$ and $np$ P-wave scattering data of the larger, more recent Granada analysis that could not be reliably extracted in the earlier Nijmegen analysis.

Having two models of the two-nucleon interaction, as well as different three-body forces, allows us to investigate how different choices of the input Hamiltonian impact the predictions of $\delta_C$ and $V_{ud}$. 
Namely, we can investigate how different approaches to regularizing short-range singularities in the interaction and different three-nucleon forces impact our calculation. 
The AV18 is a relatively hard interaction, with the central potential taking values up to ${\sim}2$ GeV~\cite{Carlson:2014vla}. 
The NV2-Ia, instead, has a softer core and the regularization of short-range singularities corresponds to a cutoff of ${\sim} 500$ MeV in momentum space~\cite{Piarulli:2016vel}. 
The different three-body fitting schemes can also have an impact on the prediction of static observables in light nuclei, as the AV18+IL7 models and the NV2+3-Ia produce accurate energy spectra of light nuclei with $A\leq 12$ compared to their counterparts~\cite{Carlson:2014vla,Piarulli:2017dwd}. 
With these four models, we are able to make an {\it ad hoc} uncertainty estimate for the predicted quantities; however, this certainly does not constitute a robust quantification of uncertainties arising from the nuclear interaction. 
Although there have been recent developments of both nuclear interactions with robustly quantified uncertainties~\cite{Bub:2024gyz,Somasundaram:2023sup,Thim:2023fnl,Svensson:2023twt} and emulators for various nuclear observables~\cite{Konig:2019adq,Wesolowski:2021cni,Odell:2023cun,Becker:2023dqe,Somasundaram:2024zse,Armstrong:2025tza}, such an analysis would be beyond the scope of the present study. 

\section{Quantum Monte Carlo methods}
\label{sec:qmc}
To obtain the many-body wave functions needed to evaluate matrix elements of the Fermi transition operator, we employ continuum quantum Monte Carlo (QMC) methods. QMC comprises a class of stochastic techniques designed to solve the Schr\"{o}dinger equation for strongly correlated many-body systems with high accuracy. Their application to the structure of finite nuclei has been covered extensively~\cite{Carlson:2014vla,Gandolfi:2020pbj}, and more recently their application to the study of electroweak processes was reviewed in Ref.~\cite{King:2024zbv}. Here, we will overview the salient features of the approaches, with emphasis on details relevant for $\delta_C$, and direct interested readers to the aforementioned review articles for additional information. 

\subsection{Variational Monte Carlo}

Continuum QMC calculations of finite nuclei typically proceed in two stages. The first is variational Monte Carlo (VMC), which is used to construct and optimize a trial many-body wave function. The goal is to produce well-correlated states $\Psi_V(J^\pi;T,T_z)$ with specific spin-parity $J^\pi$, isospin $T$, and isospin projection $T_z$ quantum numbers that have a reasonable energy spectrum and spatial extent. To achieve this, we make an ansatz for a variational trial wave function $\Psi_V$ of the form~\cite{Pudliner:1997ck},
\begin{equation}
\ket{\Psi_V} = \mathcal{S}\prod_{i<j}\left[ 1 + U_{ij} + \sum_{i<j\neq k} \widetilde{U}^{\rm TNI}_{ijk} \right] \ket{\Psi_J}\, ,\label{eq:psi.t}
\end{equation}
where $\mathcal{S}$ is the symmetrization operator, $U_{ij}$ is a two-body correlation operator, $\widetilde{U}^{\rm TNI}_{ijk}$ is a three-body correlation operator, and the Jastrow wave function $\Psi_J$ is:
\begin{equation}
   |\Psi_J\rangle = \mathcal{A}\prod_{i<j}f_c(r_{ij}) |\Phi_A(J^\pi;T T_z)\rangle \ .
\end{equation}
Here $\mathcal{A}$ is the antisymmetrization operator and the single-particle $A$-body wave function $\Phi_A(J^\pi;T T_z)$ is constructed from a sum of different spatial symmetry components with the quantum numbers of the state of interest.  
In this case, for the $0^+;1$ states in $^{10}$C and $^{10}$B, the spatial symmetry states that can be built by the coupling of four s-shell and six p-shell nucleons are, in spin-orbit labeling, $^1$S[442], $^3$P[4411], $^3$P[433], $^5$D[4321], $^3$P[4321], and $^1$S[4222], where [$n$] is the Young diagram specification~\cite{Wiringa:2006ih}.
(We note for later discussion that while all these components contribute to a $T=1$ state, the $^5$D[4321] can also be $T=0$, the $^3$P[4321] can be $T=0,2$, and the $^1$S[4222] can be $T=3$.)
The product over all pairs of the central two-body correlation $f_c(r_{ij})$ keeps nucleons apart to avoid the strong short-range repulsion of the nucleon-nucleon force. 
In practice, this correlation is broken into three subtypes that  can vary in their long-range behavior -- $f_{ss}$, $f_{pp}$, and $f_{sp}$ -- depending on whether both particles are in the s-shell or the p-shell, or one in each shell.
The antisymmetrization is carried out explicitly by summing over all partitions of the ten nucleons into s- and p-shell orbitals.

The two-body correlation operator has the structure 
\begin{equation}
   U_{ij} = \sum_{p=2,6} u_{p}(r_{ij}) O^{p}_{ij} \ ,
\end{equation}
where $O^{p}_{ij}$ are the leading spin, isospin, spin-isospin, tensor, and tensor-isospin operators in $v_{ij}$.
Perturbation theory is used to motivate the three-body correlation operator
\begin{equation}
\tilde{U}^{TNI}_{ijk} = -\epsilon \tilde{V}_{ijk}(\tilde{r}_{ij},
                     \tilde{r}_{jk}, \tilde{r}_{ki}) \ ,
\end{equation}
where $\tilde{r}=yr$, $y$ is a scaling parameter, $\epsilon$ is a small strength parameter, and $\tilde{V}_{ijk}$ includes the anticommutator part of two-pion exchange and any short-range repulsion in the three-nucleon potential.
We allow for different strengths $\epsilon$ and scaling factors $y$ for different parts of $\tilde{V}_{ijk}$.

Additional correlation operators for the CIB parts of $v_{ij}$ are included in an analogous manner by adding
\begin{equation}
   U^{CIB}_{ij} = -\omega v^{CIB}_{ij}(\tilde{r}_{ij}) \ ,
\label{eq:ucib}
\end{equation}
where $v^{CIB}_{ij}$ includes both the electromagnetic  $v_\gamma(r_{ij})$ and the $O^{p=15-18}_{ij}$ CIB parts of the nuclear interaction.
We allow for four different strengths $\omega$ for the Coulomb, magnetic moment, nuclear CD, and nuclear CSB parts of $v^{CIB}_{ij}$ as well as different scalings $y$ in $\tilde{r}$.

The variational parameters that enter the two- and three-body correlation operators are determined by minimizing the energy expectation value,
\begin{equation}
E_V = \frac{\mel{\Psi_V}{H}{\Psi_V}}{\inner{\Psi_V}{\Psi_V}}\, ,
\end{equation}
for the given Hamiltonian $H$, with the constraint that charge radii are close to expected values. This will produce a state $\Psi_V$ with an energy that is an upper bound to the true ground state energy of the system, $E_0$.
The parameters in the correlation functions are found using a simplex search over their parameter space to minimize the energy expectation value.  A final step is a small-basis diagonalization among allowed single-particle spatial symmetry components -- in this case the six $A=10,J^\pi=0^+,T=1$ spatial symmetry components -- to obtain the lowest state.

The effect of the EM and nuclear CIB parts of the interaction is to produce $^{10}$C and $^{10}$B* wave functions that are not strict isobaric analogs.   This is done in three different steps.  
First, the single-particle wave function $\Phi_A(J^\pi;T T_z)$ includes p-wave orbitals that are the solution of a nucleon in a Woods-Saxon well with an average Coulomb potential term added, which is different in $^{10}$C and $^{10}$B (see Eq.(3.20) of \cite{Pudliner:1997ck}).  
Second, the $U^{CIB}_{ij}$ correlations are turned on and their parameters optimized by calculating the energy improvement induced by their inclusion.  
Generally, these are the same in both nuclei.  Sample values are shown in Table~\ref{omega}.  
Finally, the small-basis diagonalization among spatial symmetry components is carried out separately for each nucleus, but using a common set of Monte Carlo configurations.  The variation in spatial symmetry composition is illustrated in Table~\ref{spin-space}.
The first and third steps can be made while maintaining pure isospin states with $T=1$; this type of wave function has been used in the recent QMC study of the $\delta_{\rm NS}$ correction to superallowed beta decay~\cite{King:2025fph}.
However the second step introduces $T=2,3$ components into the $^{10}$C wave function and $T=0,2,3$ components into the $^{10}$B* wave function.

\begin{table}
\begin{tabular}{|c|c| c | c | c | c |  c | c |}
\hline
Nucleus & $H$ & $\omega_C$ & $\omega_{MM}$ & $\omega_{CD}$ & $\omega_{CSB}$ & $y_{EM}$ & $y_{CIB}$ \\
\hline
$^{10}$C & AV18+IL7 & 0.02155 & 0.00092 & 0.00043 & 0.00020 & 0.55 & 0.55 \\
\hline
$^{10}$C & NV2+3-Ia & 0.02070 & 0.00160 & 0.00040 & 0.00030 & 0.55 & 0.55 \\
\hline
$^{10}$C & NV2+3-Ia$^{\star}$ & 0.02155 & 0.00092 & 0.00043 & 0.00010 & 0.55 & 0.55 \\
%$^{10}$B*&           & 0.02070 & 0.00160 & 0.00040 & 0.00020 & 0.55 & 0.55 \\
\hline
\end{tabular}
\caption{Parameters in the EM and nuclear CIB parts of the trial wave function.  The $\omega$ subscripts denote Coulomb (C), magnetic moment (MM), nuclear charge-dependent operators $O^{p=15-17}_{ij}$ (CD), and charge-symmetry-breaking $O^{p=18}_{ij}$ (CSB), while the $y$ subscripts are for the electromagnetic (EM=C+MM) and nuclear charge-independence (CIB=CD+CSB) pieces.  Parameters for AV18+UX are the same as for AV18+IL7.}
\label{omega}
\end{table}

\begin{table}
\begin{tabular}{|c|c| c | c | c | c |  c | c |}
\hline
Nucleus & $H$ & $^1$S[442] & $^3$P[4411] & $^3$P[433] & $^5$D[4321] & $^3$P[4321] & $^1$S[4222] \\
\hline
$^{10}$C & AV18+IL7 & 0.8741 & 0.4672 & 0.0921 & 0.0468 & 0.0840 &    0.0008 \\
$^{10}$B*&          & 0.8783 & 0.4581 & 0.0952 & 0.0485 & 0.0855 & $-$0.0001 \\
\hline
$^{10}$C & NV2+3-Ia & 0.8340 & 0.4986 & 0.1600 & 0.0412 & 0.1684 & $-$0.0144 \\
$^{10}$B*&          & 0.8404 & 0.4844 & 0.1661 & 0.0463 & 0.1706 & $-$0.0168 \\
\hline
$^{10}$C & NV2+3-Ia$^{\star}$& 0.8556 & 0.4810 & 0.1408 & 0.0339 & 0.1253 & $-$0.0052 \\
$^{10}$B*&          & 0.8579 & 0.4731 & 0.1486 & 0.0399 & 0.1282 & $-$0.0045 \\
\hline
\end{tabular}
\caption{Spatial symmetry components in $^{10}$C(0$^+$) and $^{10}$B*(0$^+$) states for different Hamiltonians.  Amplitudes for AV18+UX are the same as for AV18+IL7.}
\label{spin-space}
\end{table}

\subsection{Green's function Monte Carlo}

One can improve upon the best variational state $\Psi_V$ with the Green's Function Monte Carlo (GFMC) method. The method leverages the fact that the real time Schr\"{o}dinger Equation may be recast as a diffusion equation in imaginary time $\tau$. Noting that one may expand any state in a complete orthonormal basis, we could imagine $\Psi_V$ as being a linear combination of the true eigenstates $\Psi_n$ of $H$,
\begin{equation}
\ket{\Psi_V} = \sum_{n=0}^{\infty} c_n\ket{\Psi_n}\, .
\end{equation}
While we do not know the states $\Psi_n$, we can project out the true ground state by propagation in $\tau$,
\begin{equation}
\lim_{\tau\to\infty} e^{-(H-E_0)\tau}\ket{\Psi_V} \propto c_0|\Psi_0\rangle\, .
\end{equation}
In practice, propagation is performed in $n$ small steps in imaginary time $\Delta\tau$,
\begin{equation}
\ket{\Psi(\tau)} = \left[ e^{-(H-E_0)\Delta\tau}\right]^n\ket{\Psi_V}\, ,
\end{equation}
until spurious contamination is removed from the wave function and convergence is reached. When performing the propagation in imaginary time, one typically exponeniates only the interaction reprojected onto the the first eight operators of Eq.~\eqref{eq:ops},
\begin{eqnarray}
  O^{p=1,8}_{ij} & = & [1, \sigma_{i}\cdot\sigma_{j}, S_{ij},
    {\bf L\cdot S}] \otimes [1,\tau_{i}\cdot\tau_{j}] \ .
\end{eqnarray}
Here however, in this work we include both CSB and CD terms as well as the full long-range electromagnetic interaction in the propagator. 
%\begin{equation}
%    \mathcal{O}^{p=1,\ldots,8}_{ij} = [1, \bsigma^{(i)} \cdot\bsigma^{(j)},S^{(ij)}(\mathbf{\hat{r}})),\LL_{ij}\cdot\mathbf{S}_{ij}] \otimes [1,\btau^{(i)}\cdot\btau^{(j)}]\, ,
%\end{equation}
%where $S_{ij} = (\bsigma^{(i)} +\bsigma^{(j)})/2$~\cite{Pieper:2001mp}. 
The reprojected potential is generated such that the $S$- and $P$-wave scattering properties, as well as deuteron properties, are preserved. We note that in these calculations, it is typical to adjust the three-nucleon force in order to minimize the difference between the expectation value of the reprojected Hamiltonian $\avg{H'}$ and the full Hamiltonian $\avg{H}$. The difference $\avg{H'-H}$ is then treated perturbatively. In this work, we adjust the three-nucleon potentials consistent with what has previously been done for energy calculations using the nuclear Hamilitonians that we employ~\cite{Pudliner:1997ck,Pieper:2001mp,Piarulli:2017dwd}. The imaginary time propagations were performed using ${\sim} 160,000$ walkers out to an imaginary time $\tau = 0.64$ MeV$^{-1}$ for the AV18+UX and AV18+IL7 Hamiltonians, and $\tau = 0.30$ to 0.42 MeV$^{-1}$ for the NV2+3 interacations; these are sufficient to achieve convergence in energies and observables.

While one would like to compute diagonal matrix elements of the form,
\begin{equation}
    \avg{\mathcal{O}(\tau)} = \frac{\mel{\Psi(\tau)}{\mathcal{O}}{\Psi(\tau)}}{\inner{\Psi(\tau)}{\Psi(\tau)}}\, ,
\end{equation}
this would require an imaginary time propagation for each matrix element under study, and would be computationally prohibitive. To circumvent this limitation and perform an analysis of each contribution to the Fermi transition operator, we compute mixed estimates. If we take $\Psi(\tau) = \Psi_V + \delta \Psi$, where $\delta \Psi$ is a small correction, then we may write~\cite{Carlson:2014vla,Gandolfi:2020pbj},
\begin{equation}
\avg{\mathcal{O}(\tau)} \approx 2 \frac{\mel{\Psi(\tau)}{\mathcal{O}}{\Psi_V}}{\inner{\Psi(\tau)}{\Psi_V}} - \frac{\mel{\Psi_V}{\mathcal{O}}{\Psi_V}}{\inner{\Psi_V}{\Psi_V}}\, . \label{eq:mixed.diag}
\end{equation}

While the above applies for diagonal matrix elements, for $\beta$ decay, one must propagate the matrix element,
\begin{equation}
\avg{\mathcal{O}(\tau)}_{i\to f } = \frac{\mel{\Psi_f(\tau)}{\mathcal{O}}{\Psi_i(\tau)}}{\sqrt{\inner{\Psi_f(\tau)}{\Psi_f(\tau)}}\sqrt{\inner{\Psi_i(\tau)}{\Psi_i(\tau)}}}\, ,
\end{equation}
where $\Psi_{i(f)}(\tau)$ is the initial (final) state involved in the transition at imaginary time $\tau$. The mixed estimate was first generalized for off-diagonal transitions in Ref.~\cite{Pervin:2007sc}, and the resultant expression that we use in this work is,
\begin{eqnarray}\nonumber
\avg{\mathcal{O}(\tau)}_{i \to f} &\approx& \sqrt{\frac{\inner{\Psi_f}{\Psi_f}}{\inner{\Psi_i}{\Psi_i}}}\frac{\mel{\Psi_f(\tau)}{\mathcal{O}}{\Psi_i}}{\inner{\Psi_f(\tau)}{\Psi_f}} + \sqrt{\frac{\inner{\Psi_i}{\Psi_i}}{\inner{\Psi_f}{\Psi_f}}}\frac{\mel{\Psi_i(\tau)}{\mathcal{O}^{\dagger}}{\Psi_f}}{\inner{\Psi_i(\tau)}{\Psi_i}}\\ 
&&- \frac{\mel{\Psi_f}{\mathcal{O}}{\Psi_i}}{\sqrt{\inner{\Psi_f}{\Psi_f}}\sqrt{\inner{\Psi_i}{\Psi_i}}}\, . \label{eq:mixed.off.diag}
\end{eqnarray}
The quantity in Eq.~\eqref{eq:mixed.off.diag} is what is propagated in imaginary time, and then averaged once convergence is reached. 

\section{Results}
\label{sec:results}

\begin{figure}
    \centering
    \includegraphics[width=\linewidth]{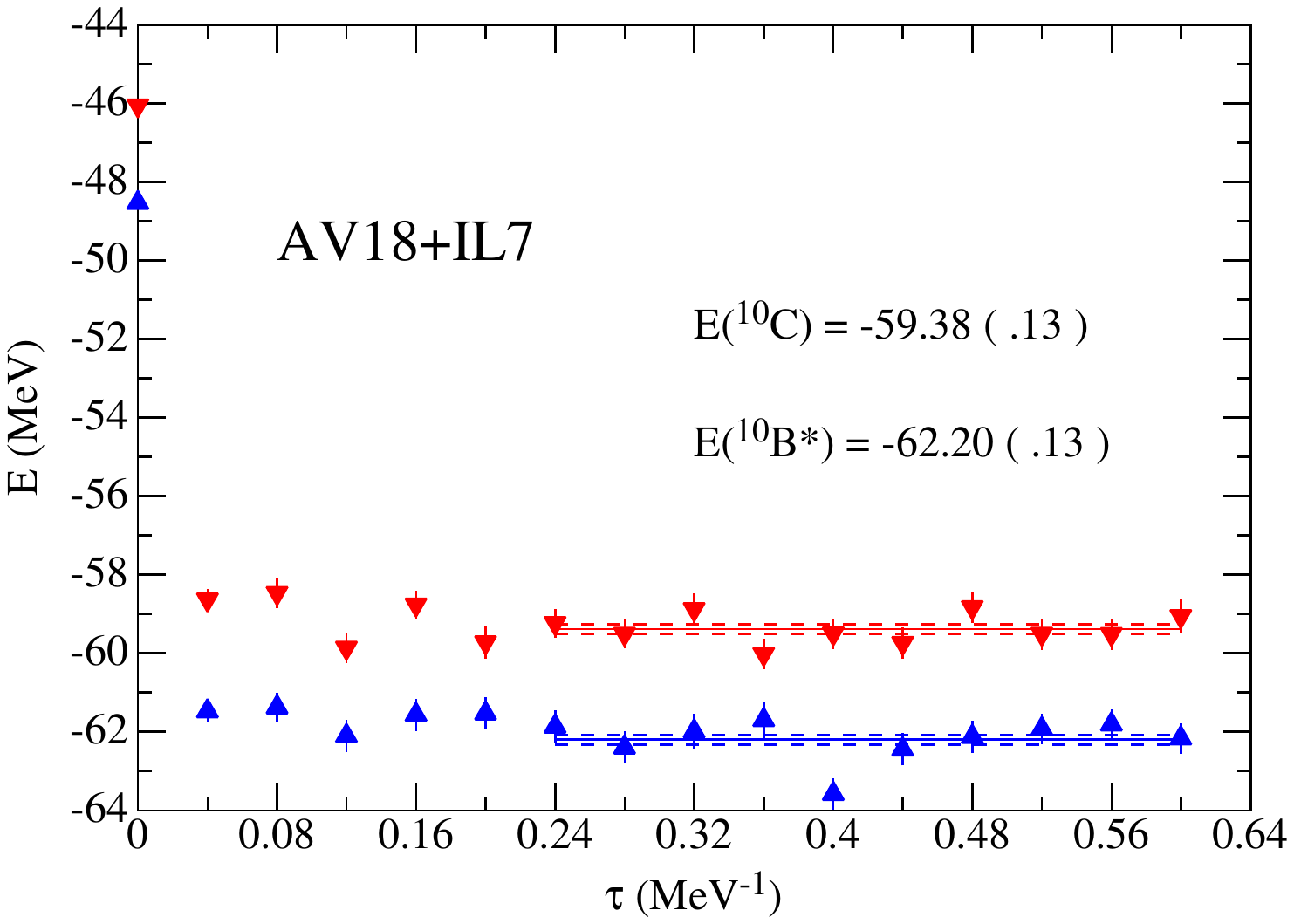}
    \caption{GFMC propagation for the energy of $^{10}$C and $^{10}$B* for the AV18+IL7 Hamiltonian as a function of the imaginary time $\tau$.  
    Red down triangles with error bars are for $^{10}$C while blue up triangles are for $^{10}$B*; solid lines show the average over $\tau = 0.24-0.60$ MeV$^{-1}$ and dashed lines are the corresponding error estimate.
    }
    \label{fig:energy_tau}
\end{figure}

The energies for $^{10}$C and $^{10}$B* calculated in both VMC and GFMC are shown in Table~\ref{tab:energy} for four Hamiltonians: AV18+UX, AV18+IL7, NV2+3-Ia, and NV2+3-Ia$^{\star}$.  An example of the energy evolution with $\tau$ is shown for AV18+IL7 in Fig.~\ref{fig:energy_tau}.  The GFMC calculation generally improves on the VMC estimate by 11 to 13 MeV, which is a substantial fraction of the total binding.
However, the binding energy itself is the result of a large cancelation between the kinetic, two-body potential, and three-body potential terms as illustrated in Table~\ref{tab:carbon} for the case of $^{10}$C, where we show the VMC and mixed estimates from GFMC.  The values for $^{10}$B* are very similar, except the electromagnetic contribution is smaller at 5.2-5.5 MeV.
In general, the VMC calculations get $\approx 95\%$ of the kinetic and two-body potential contributions, but only $\approx 75\%$ of the three-body potential contribution.

The experimental energy for the ground state of $^{10}$C is -60.32 MeV and -63.01 MeV for the first $0^+$ excited state of $^{10}$B; the four Hamiltonians bracket these values, with AV18+IL7 underbound by $\sim 1$ MeV and NV2+3-Ia overbound by $\sim 2.5$ MeV being the closest.  For the point nucleon rms radii we show the VMC and GFMC values, which agree within $\pm5\%$; the GFMC propagation increases the difference between proton and neutron radii in $^{10}$C by about 0.1 fm over VMC, but only about 0.01 fm in $^{10}$B$^*$.
There are currently no experimental values for these radii; however, we can compare with the recommended values from Ref.~\cite{OHAYON2025101732}, which use a mirror shift formula for the ground state radius of $^{10}$C and a semi-empirical formula for isotriplets to deduce the excited state radius of $^{10}$B$^*$. 
The recommended values for $R_{ch}$ are 2.64 fm and 2.54 fm for $^{10}$C and $^{10}$B$^*$, respectively. The AV18+IL7 provides values with ${\sim}$2.6\% (${\sim}$4.0\%) agreement for $^{10}$C ($^{10}$B$^*$). The chiral models NV2+3-Ia and NV2+3-Ia* agree with the recommendations at the ${\sim}$4\% to ${\sim}$5\% level, while the AV18+UX deviates from the recommendation by ${\sim}$7.6\% (${\sim}$6.3\%) for $^{10}$C ($^{10}$B$^*$). Again, the four models bracket the recommended values, with model Ia providing smaller radii and the remaining models providing larger values. 

\begin{table}
\begin{tabular}{|c|c| c | c | c | c | c | c | c | c |}
\hline
Nucleus & $H$ &
\multicolumn{2}{c|} {$E$ (MeV)} &
{\# configs} &
\multicolumn{2}{c|} {$r_p$ (fm)} &
{$R_{ch}$ (fm)} &
\multicolumn{2}{c|} {$r_n$ (fm)} \\
\hline
& & VMC & GFMC & GFMC & VMC & GFMC & GFMC & VMC & GFMC \\
\hline
$^{10}$C & AV18+UX  & -43.19(5) & -55.35(12) & 1.4M+ & 2.49 & 2.72 & 2.84 & 2.28 & 2.38 \\
$^{10}$B*&          & -45.58(4) & -57.75(12) & 1.4M+ & 2.40 & 2.56 & 2.69 & 2.39 & 2.54 \\\hline
$^{10}$C & AV18+IL7 & -46.06(5) & -59.38(13) & 1.4M+ & 2.49 & 2.57 & 2.71 & 2.28 & 2.28 \\
$^{10}$B*&          & -48.53(5) & -62.20(14) & 1.4M+ & 2.40 & 2.44 & 2.57 & 2.39 & 2.42 \\
\hline
$^{10}$C & NV2+3-Ia & -50.35(5) & -62.27(30) & 2.2M+ & 2.43 & 2.37 & 2.52 & 2.25 & 2.12 \\
$^{10}$B*&          & -53.27(6) & -66.04(28) & 2.2M+ & 2.35 & 2.29 & 2.43 & 2.34 & 2.27 \\
\hline
$^{10}$C & NV2+3-Ia$^{\star}$& -46.36(5) & -56.13(31) & 1.4M+ & 2.62 & 2.64 & 2.77 & 2.37 & 2.30 \\
$^{10}$B*&          & -48.80(5) & -59.42(27) & 1.4M+ & 2.51 & 2.50 & 2.63 & 2.49 & 2.48 \\
\hline
\end{tabular}
\caption{Energies (in MeV) and point proton, charge, and point neutron radii (in fm) of $^{10}$C(0$^+$) and $^{10}$B*(0$^+$) states for different Hamiltonians.  MC errors are shown in parentheses except when less than 1 in the last place.}
\label{tab:energy}
\end{table}

\begin{table}
\begin{tabular}{|c| r | r | r | r | r | r |  r | r |}
\hline
& 
\multicolumn{2}{c|} {AV18+UX} &
\multicolumn{2}{c|} {AV18+IL7} &
\multicolumn{2}{c|} {NV2+3-Ia} &
\multicolumn{2}{c|} {NV2+3-Ia$^{\star}$} \\
\hline
$^{10}$C   & {\rm VMC}&{\rm mixed}& {\rm VMC}&{\rm mixed}& {\rm VMC}&{\rm mixed}& {\rm VMC}&{\rm mixed}\\
\hline
$T$        & $ 288.8$ & $ 283.7$ & $ 288.9$ & $ 294.0$ & $ 241.3$ & $ 259.6$ & $ 211.3$ & $ 224.5$ \\
$v^N_{ij}$ & $-330.4$ & $-335.3$ & $-330.5$ & $-345.2$ & $-282.1$ & $-307.6$ & $-257.2$ & $-278.5$ \\
$v^{EM}_{ij}$ & $7.6$ &    $7.4$ &   $ 7.6$ & $   7.6$ & $   8.0$ & $   8.0$ & $   7.5$ & $   7.5$ \\
$V_{ijk}$  &   $-9.2$ &  $-11.4$ & $ -12.1$ & $ -16.1$ & $ -17.5$ & $ -23.0$ & $  -7.9$ & $ -10.1$ \\
\hline
$E$        &  $-43.2$ &  $-55.5$ & $ -46.1$ & $ -59.7$ & $ -50.3$ & $ -63.0$ & $ -46.3$ & $ -56.6$ \\
\hline
\end{tabular}
\caption{Energy breakdown (in MeV) of $^{10}$C(0$^+$) ground state in VMC and GFMC calculations for different Hamiltonians.  This is just to show relative sizes of different terms, so MC errors are not shown.}
\label{tab:carbon}
\end{table}

\begin{table}
\begin{tabular}{|c|c| c | c | c | c |  c | c |}
\hline
Nucleus & $H$ &
\multicolumn{2}{c|} {$T=0$} &
\multicolumn{2}{c|} {$T=2$} &
\multicolumn{2}{c|} {$T=3$} \\
\hline
& & VMC & mixed & VMC & mixed & VMC & mixed \\
\hline
$^{10}$C & AV18+UX  & ---    & ---        & 0.0192 & 0.0120(2) & 0.0032 & 0.0029(1)\\
$^{10}$B*&          & 0.0131 & 0.0270(67) & 0.0133 & 0.0078(1) & 0.0048 & 0.0043(1)\\
\hline
$^{10}$C & AV18+IL7 & ---    & ---        & 0.0192 & 0.0130(3) & 0.0032 & 0.0031(1)\\
$^{10}$B*&          & 0.0131 & 0.0230(91) & 0.0133 & 0.0084(2) & 0.0048 & 0.0046(1)\\
\hline
$^{10}$C & NV2+3-Ia & ---    & ---        & 0.0261 & 0.0217(10)& 0.0046 & 0.0085(8)\\
$^{10}$B*&          & 0.0136 & 0.0201(51) & 0.0173 & 0.0080(5) & 0.0069 & 0.0132(17)\\
\hline
$^{10}$C & NV2+3-Ia$^{\star}$& ---    & ---        & 0.0227 & 0.0176(4) & 0.0036 & 0.0063(1)\\
$^{10}$B*&          & 0.0147 & 0.0246(21) & 0.0181 & 0.0085(3) & 0.0066 & 0.0107(2)\\
\hline
\end{tabular}
\caption{Percentage CIB isospin components in $^{10}$C(0$^+$) and $^{10}$B(0$^+$) states for different Hamiltonians.  MC errors are shown in parentheses except when less than 1 in the last place.}
\label{tab:isospin}
\end{table}

\begin{figure*}[t]
\centering

\begin{minipage}{0.49\textwidth}
\centering
\includegraphics[width=\linewidth,
trim=50 20 50 40, clip]{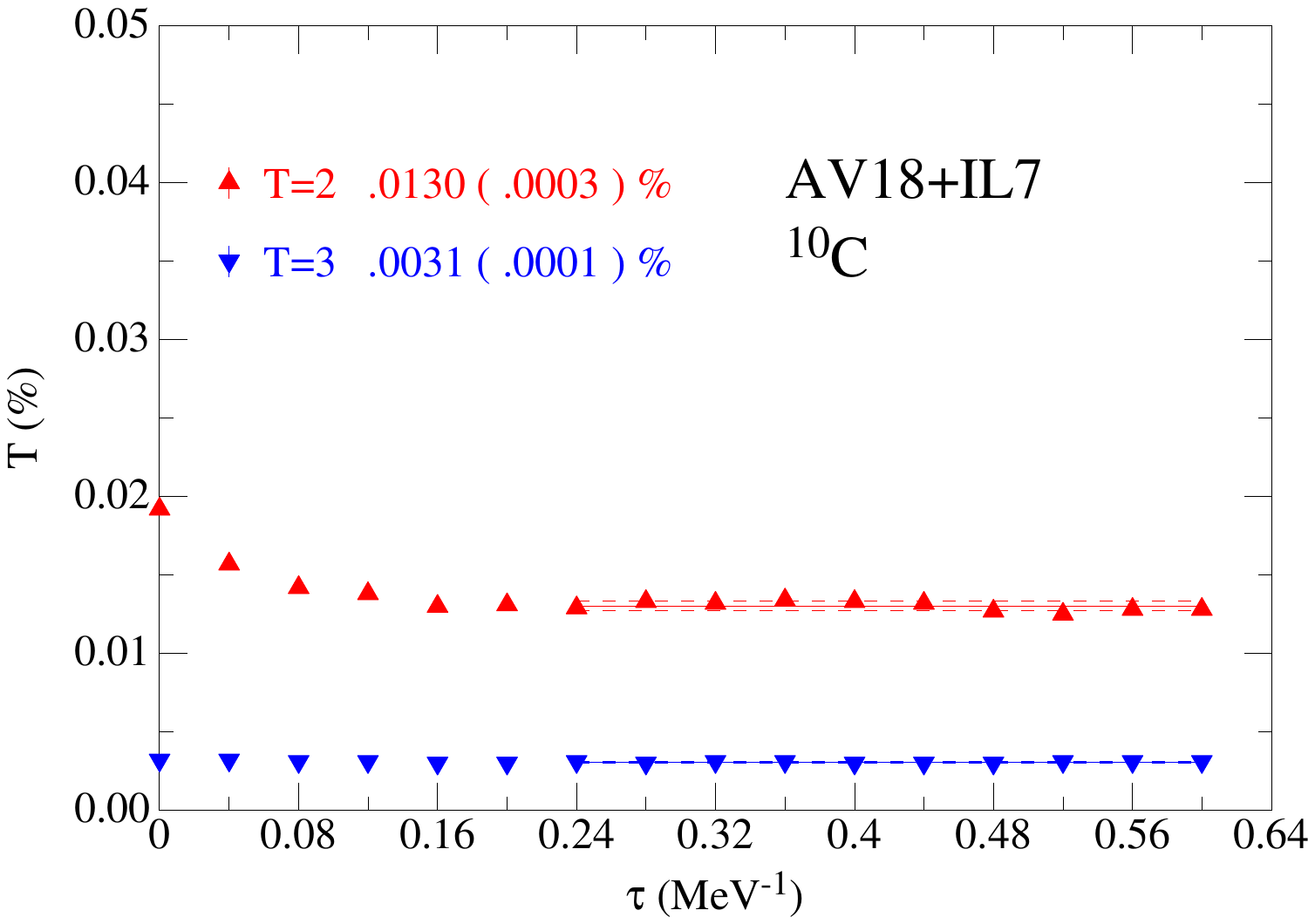}\\
\textbf{(a)}
\end{minipage}
\hfill
\begin{minipage}{0.49\textwidth}
\centering
\includegraphics[width=\linewidth,
trim=50 20 50 40, clip]{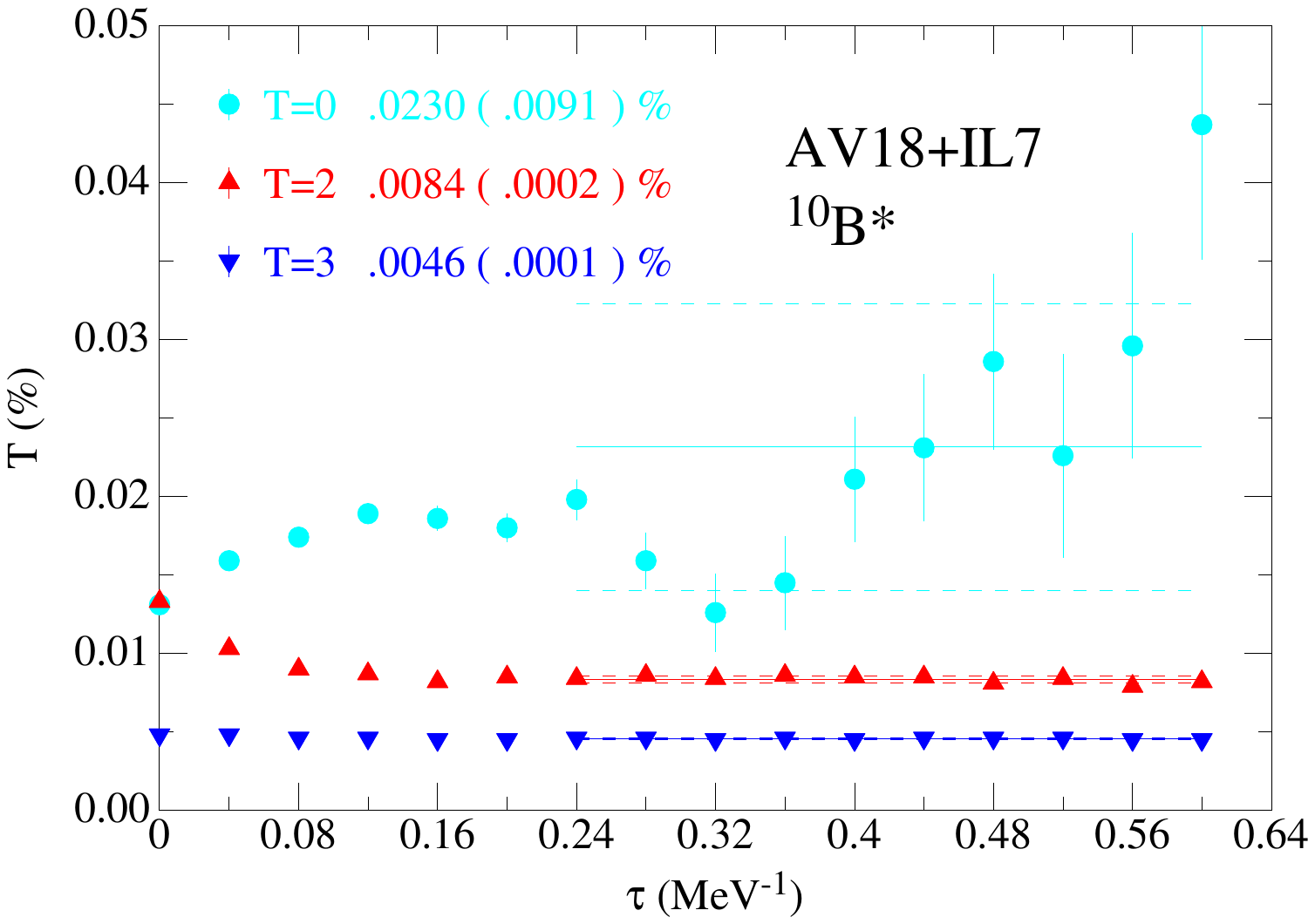}\\
\textbf{(b)}
\end{minipage}

\caption{
GFMC propagation of the $T \ne 1$ isospin content as a function of imaginary time $\tau$ for the AV18+IL7 Hamiltonian. 
Panel (a) shows $^{10}$C, where red and blue dots correspond to $T=2$ and $T=3$, respectively.
Panel (b) shows $^{10}$B, where green dots correspond to $T=0$, and red and blue to $T=2$ and $T=3$, respectively. 
In both panels, solid lines indicate averages over $\tau = 0.24$--$0.60$ MeV$^{-1}$, while dashed lines represent the corresponding uncertainty estimates.
}
\label{fig:iso_combined}

\end{figure*}

The VMC and GFMC mixed estimates of the $T\ne1$ isospin projections in the $^{10}$C and $^{10}$B* states are shown in Table~\ref{tab:isospin}.  
An example of the GFMC propagation for these projections is shown in Fig.~\ref{fig:iso_combined} for the AV18+IL7 Hamiltonian.  
While the $T=2,3$ content is very stable for both nuclear states, the $T=0$ content in $^{10}$B* is not well converged and consequently has a large error bar in Table~\ref{tab:isospin}.
The $T=0$ expectation values for NV2+3-Ia and -Ia$^{\star}$ are more stable, but still have much larger error bars than the $T=2,3$ components.  The slower convergence of the $T=0$ component may be related to its structure in the wave function. In particular, the $T=0$ admixture arises from mixing with nearby states involving different spatial symmetry components and may be generated through cancellations among larger contributions. As a result, its expectation value could be obtained as a small residual, making it more sensitive to statistical fluctuations in the Monte Carlo sampling. In contrast, the $T=2$ and $T=3$ components are more cleanly separated in symmetry space and tend to exhibit a more stable behavior during the GFMC propagation.
Moreover, VMC calculations indicate that the nearest $0^+$, $T=0$ state in $^{10}$B (built upon the $^3$P[4321] and $^5$D[4321] spatial symmetry states) is $\approx 16$ MeV higher in energy than the $0^+$, $T=1$ state, so isospin mixing between these states should be extremely small.
The $0^+$, $T=2$ state built on $^3$P[4321] should be $\approx 21$ MeV higher, and the $0^+$, $T=3$ state built on $^1$S[4222] (the isobaric analog to $^{10}$He ground state) should be $\approx 35$ MeV higher.
Naively, one might think the isospin mixing to be smaller for states of higher excitation, so we expect $T=2 > T=3$ in $^{10}$C, which is the case, and $T=0 > T=2 > T=3$ in $^{10}$B*, but this is not true for the NV2+3 potentials.  This indicates that the relative size of the isospin admixtures is not determined solely by energy splittings, but also depends on the detailed structure of the interaction and the coupling matrix elements between states of different isospin. In particular, the different balance of EM and CIB terms in the NV2+3 interactions can modify the mixing pattern, leading to deviations from the naive ordering.

%\begin{figure}
%    \centering
%    \includegraphics[width=\linewidth]{fig_isoc_il7.pdf}
%    \caption{GFMC propagation of the $T\ne1$ isospin content of $^{10}$C for the AV18+IL7 Hamiltonian as a function of the imaginary time $\tau$.  Red (blue) dots with error bars are for $T=2$ ($T=3$) while solid lines show the average over $\tau = 0.24-0.60$ MeV$^{-1}$ and dashed lines are the corresponding error estimate.}
%    \label{fig:isoc_tau}
%\end{figure}

%\begin{figure}
%    \centering
%    \includegraphics[width=\linewidth]{fig_isob_il7.pdf}
%    \caption{GFMC propagation of the $T\ne1$ isospin content of $^{10}$B for the AV18+IL7 Hamiltonian as a function of the imaginary time $\tau$.  Green dots with error bars are for $T=0$, while red and blue are for $T=2$ and $T=3$; solid lines show the averages over $\tau = 0.24-0.60$ MeV$^{-1}$ and dashed lines are the corresponding error estimate.}
%    \label{fig:isob_tau}
%\end{figure}

%\begin{figure}[t]
%\centering
%\includegraphics[width=0.5\linewidth, trim=40 30 40 30, clip]{fig_isoc_il7.pdf}%
%\includegraphics[width=0.5\linewidth, trim=40 30 40 30, clip]{fig_isob_il7.pdf}
%\caption{GFMC propagation of the $T\ne1$ isospin content from the AV18+IL7 Hamiltonian as a function of the imaginary time $\tau$ for $^{10}$C (left) and $^{10}$B* (right). Green dots with error bars are for $T=0$, red down-triangles for $T=2$, and blue up-triangles for $T=3$; solid lines show the average over $\tau = 0.24$--$0.60$ MeV$^{-1}$ and dashed lines are the corresponding error estimate.}
%\label{fig:iso_tau_av18}
%\end{figure}

\begin{figure*}[t!]
\centering

% --- Top row ---
\begin{minipage}{0.5\textwidth}
\centering
\includegraphics[width=\linewidth, trim=50 20 50 40, clip]{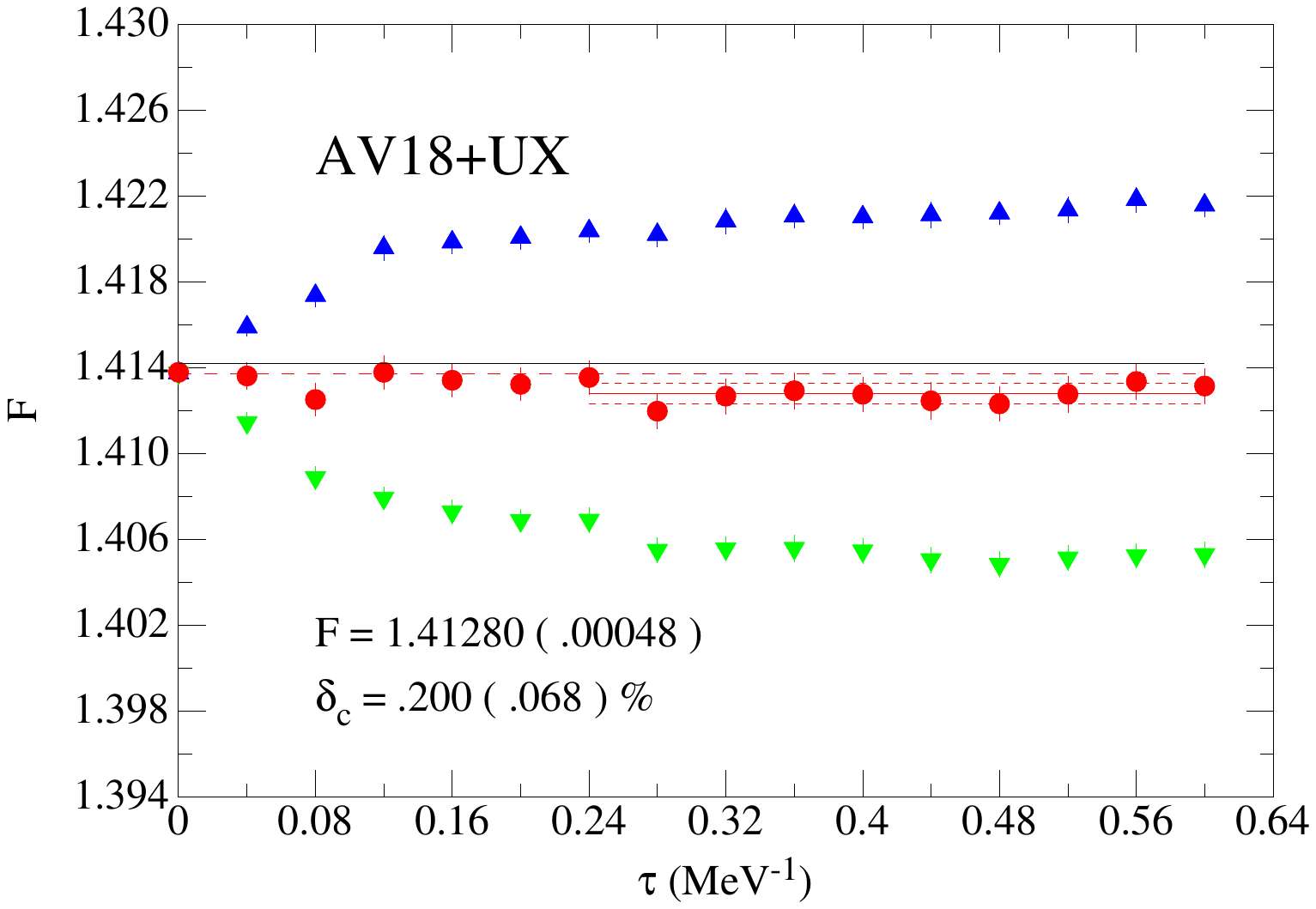}\\[-0.2em]
\textbf{(a)}
\end{minipage}%
\begin{minipage}{0.5\textwidth}
\centering
\includegraphics[width=\linewidth, trim=50 20 50 40, clip]{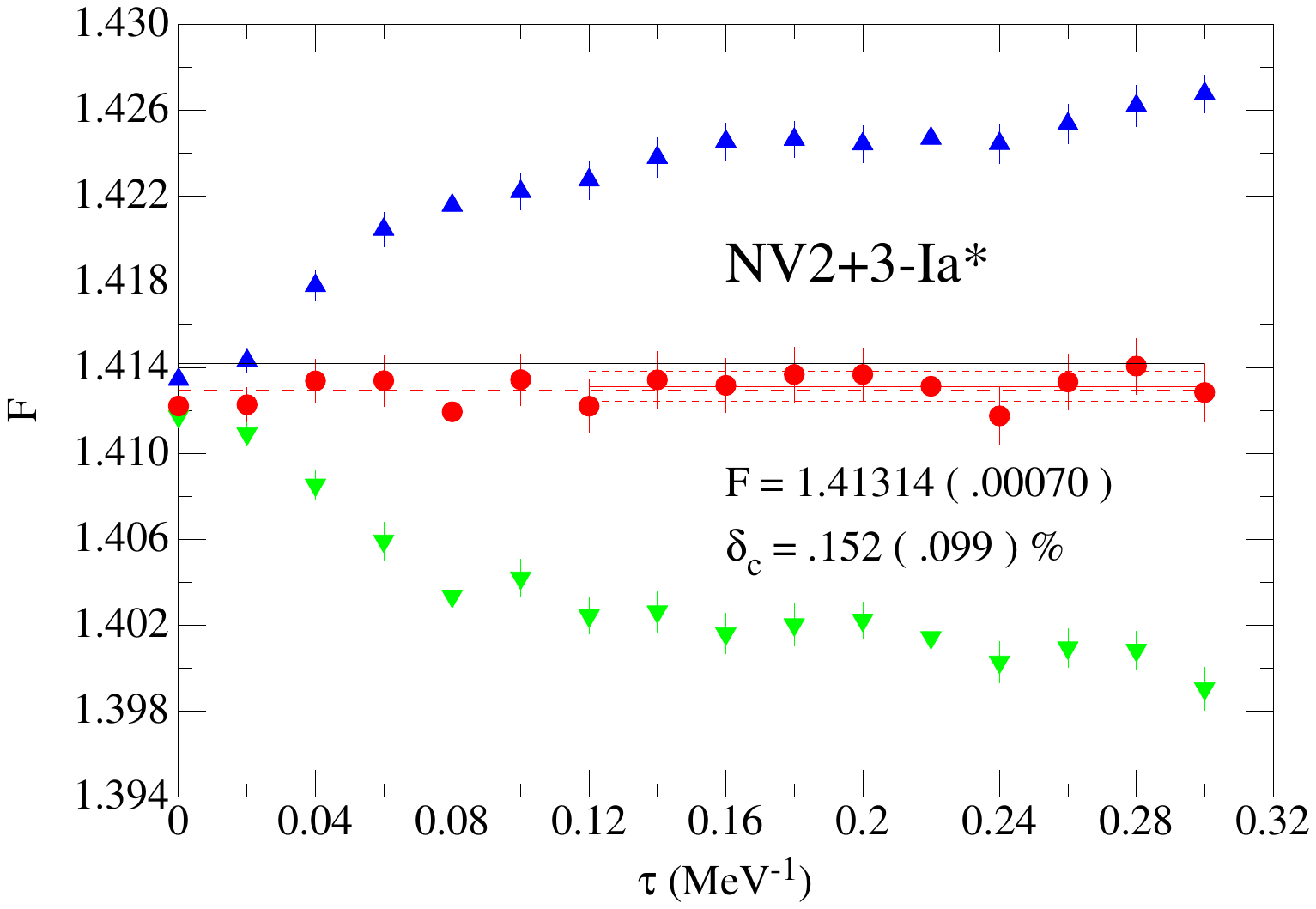}\\[-0.2em]
\textbf{(b)}
\end{minipage}

\vspace{0.2em}

% --- Bottom ---
\begin{minipage}{0.5\textwidth}
\centering
\includegraphics[width=\linewidth, trim=50 20 50 40, clip]{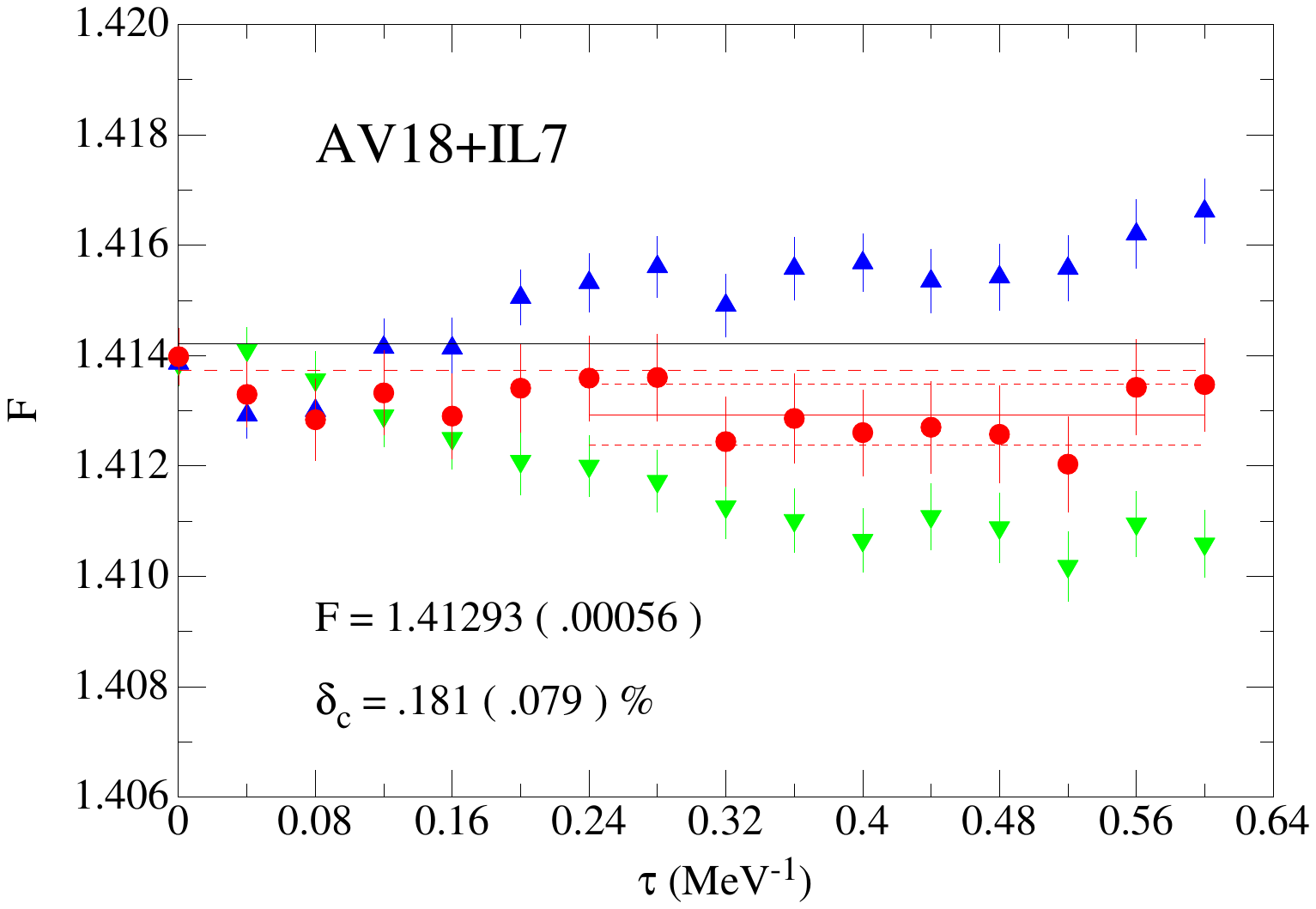}\\[-0.2em]
\textbf{(c)}
\end{minipage}%
\begin{minipage}{0.5\textwidth}
\centering
\includegraphics[width=\linewidth, trim=50 20 50 40, clip]{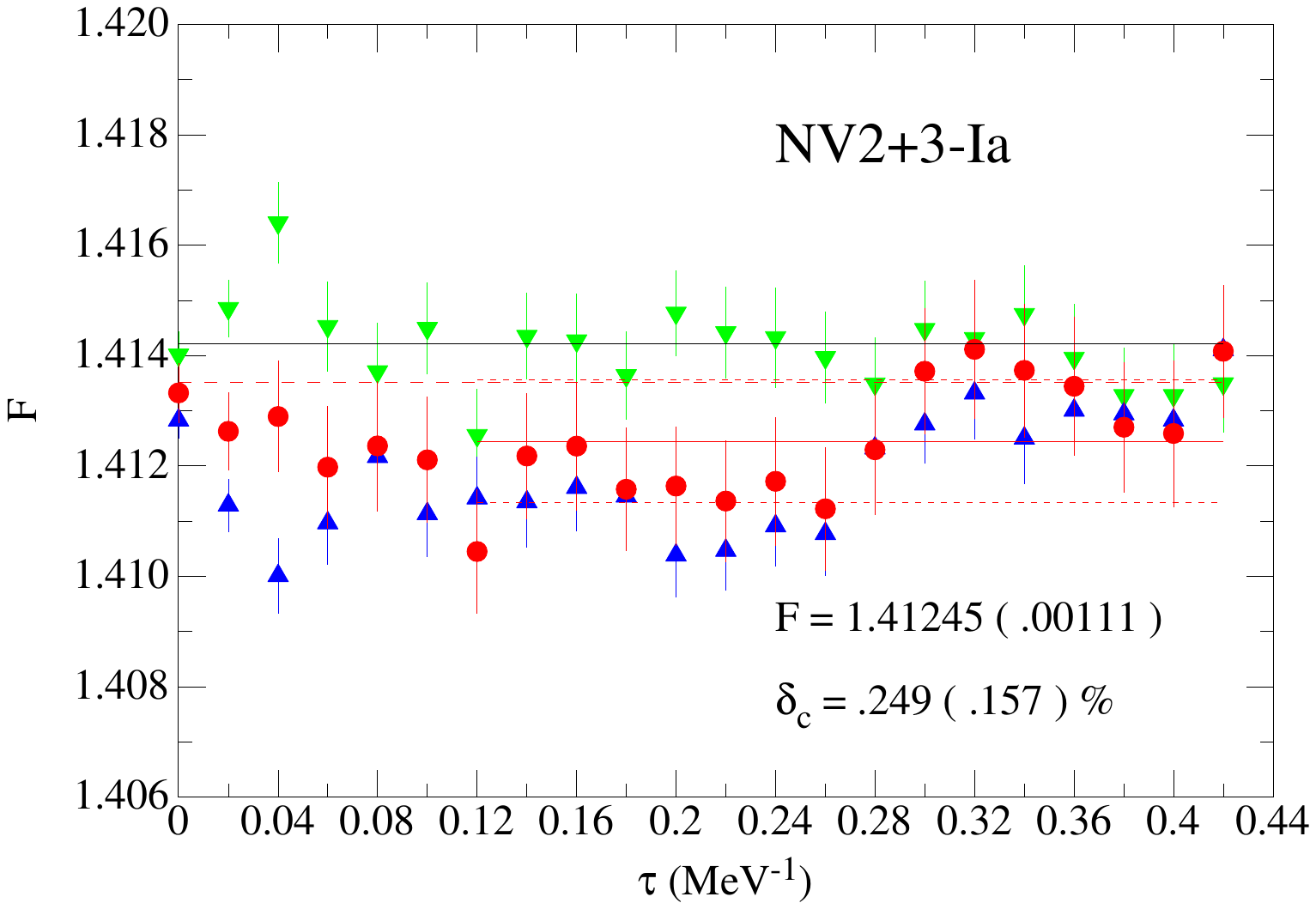}\\[-0.2em]
\textbf{(d)}
\end{minipage}

\caption{
GFMC propagation of the Fermi matrix element as a function of the imaginary time $\tau$ for (in order of increasing binding energy) (a) AV18+UX, (b) NV2+3-Ia$^{\star}$, (c) AV18+IL7, and (d) NV2+3-Ia Hamiltonians. The black solid line indicates the canonical $\sqrt{2}$ value, while the red dashed line shows the VMC estimate. Blue up triangles are the normalized GFMC mixed matrix elements $\mel{\Psi_{B^\ast}(\tau)}{F}{\Psi_C}$, green down triangles are $\mel{\Psi_C(\tau)}{F^{\dagger}}{\Psi_{B^\ast}}$, and red circles denote the GFMC extrapolation of Eq.~(\ref{eq:mixed.off.diag}). The red solid line represents the average over $\tau$, with dotted lines indicating the corresponding Monte Carlo uncertainty; the averaging ranges are $\tau = 0.24$--$0.60$ MeV$^{-1}$ for (a) and (c), $0.12$--$0.30$ MeV$^{-1}$ for (b), and $0.12$--$0.42$ MeV$^{-1}$ for (d).
}
\label{fig:ftau_combined}

\end{figure*}

The GFMC propagation in $\tau$ of the Fermi matrix element 
$\mel{^{10}{\rm B^*(0^+;1)}}{F}{^{10}{\rm C}(0^+;1)}$ is shown in Fig.~\ref{fig:ftau_combined} for the AV18+UX (a), NV2+3-Ia$^{\star}$ (b),  AV18+IL7 (c), and NV2+3-Ia (d) Hamiltonians.
In these figures, the black solid line is the canonical $\sqrt2$, corresponding to the exact Fermi matrix element for a pure $T=1$ isobaric analog transition, and the red dashed line is the VMC estimate.  Blue up triangles are the normalized GFMC mixed matrix elements $\mel{\Psi_{B^\ast}(\tau)}{F}{\Psi_C}$, green down triangles are the $\mel{\Psi_C(\tau)}{F^{\dagger}}{\Psi_{B^\ast}}$, and red circles are the GFMC extrapolation of Eq.~(\ref{eq:mixed.off.diag}).  The red solid line is the average over larger $\tau$ values (the range differs from case to case) and the dotted lines above and below give the Monte Carlo error estimate.

%%%%%%%%%%%%%%%%%%%%%%%%%%%%%

For AV18+UX and AV18+IL7, the mixed matrix element when $^{10}$B* is propagated exceeds $\sqrt2$ and gets progressively larger with $\tau$, while the opposite is true when $^{10}$C is propagated.  Despite this, the extrapolated GFMC result is reasonably constant, although with a significant error bar.
For NV2+3-Ia, the mixed matrix elements are reversed in their order, and much flatter with $\tau$.  The extrapolated GFMC matrix element is noisier with a larger error bar.  In these three cases, the GFMC result is a significantly larger deviation from $\sqrt2$ than the initial VMC estimate.
For NV2+3-Ia$^{\star}$, the mixed matrix elements show a larger deviation from $\sqrt2$ but again a very flat extrapolated result, although again with a large error bar.  In this case, the GFMC result deviates slightly less from $\sqrt2$ than the initial VMC estimate.

\begin{table}
\begin{tabular}{|c| l | l | l | l | l | l |  l | l |}
\hline
&
\multicolumn{2}{c|} {\rm AV18+UX} & 
\multicolumn{2}{c|} {\rm AV18+IL7} &
\multicolumn{2}{c|} {\rm NV2+3-Ia} &
\multicolumn{2}{c|} {\rm NV2+3-Ia$^{\star}$} \\
\hline
           & {\rm VMC}&{\rm GFMC}& {\rm VMC}&{\rm GFMC}& {\rm VMC}&{\rm GFMC}& {\rm VMC}&{\rm GFMC}\\
\hline
$F$        & $1.413729(2)$ & $1.41280(48)$& $1.413729(2)$ & $1.41293(56)$ & $1.413518(4)$ & $1.41245(111)$ & $1.412978(3)$ & $ 1.41314(70)$ \\
$\delta_C$ & $0.0685(3)$   & $0.200(68)$ & $0.0685(3)$   & $0.181(79)$ & $0.0983(6)$   & $0.249(157)$ & $0.1747(4)$   & $0.152(99)$ \\
\hline
\end{tabular}
\caption{VMC and GFMC matrix elements for $\mel{^{10}{\rm B^*(0^+;1)}}{F}{^{10}{\rm C}(0^+;1)}$ and derived values for $\delta_C$ in \%.}
\label{tab:deltac}
\end{table}

The values of the Fermi matrix element and the corresponding $\delta_C$ are summarized in Table~\ref{tab:deltac}. The GFMC results give $F = 1.41280(48)$ and $\delta_C = 0.200(68)\%$ for AV18+UX (averaged over $\tau = 0.24$--$0.60$ MeV$^{-1}$), $F = 1.41293(56)$ and $\delta_C = 0.181(79)\%$ for AV18+IL7 (averaged over $\tau = 0.24$--$0.60$ MeV$^{-1}$), $F = 1.41245(111)$ and $\delta_C = 0.249(157)\%$ for NV2+3-Ia (averaged over $\tau = 0.12$--$0.42$ MeV$^{-1}$), and $F = 1.41314(70)$ and $\delta_C = 0.152(99)\%$ for NV2+3-Ia$^{\star}$ (averaged over $\tau = 0.12$--$0.30$ MeV$^{-1}$). This indicates an ordering in which $\delta_C$ is largest for NV2+3-Ia, followed by AV18+UX, then AV18+IL7, and smallest for NV2+3-Ia$^{\star}$. The corresponding $\delta_C$ values carry large relative (fractional) uncertainties of approximately $34\%$, $44\%$, $63\%$, and $65\%$, respectively. The larger relative uncertainties obtained with the NV2+3 interactions are also related to the shorter imaginary-time propagation intervals used for the averaging, which provide a more limited plateau region for the extraction of $\delta_C$.

%This sensitivity is further enhanced by cancellations inherent in the mixed-estimator procedure. In particular, while the NV2+3 interactions exhibit a flatter behavior in $\tau$ for the Fermi matrix element, the extraction of $\delta_C$ involves differences between closely spaced contributions, which increases the impact of statistical fluctuations. In contrast, the stronger $\tau$ dependence of the Fermi matrix element observed for AV18+IL7 provides a more constrained extrapolation.

%Within these uncertainties, the observed ordering of $\delta_C$ follows the trends in binding energies and radii produced by these Hamiltonians, and in particular correlates with the expectation value of the electromagnetic (predominantly Coulomb) potential $v^{EM}_{ij}$ reported in Table~\ref{tab:carbon}. This is consistent with the Coulomb interaction being the dominant source of isospin mixing in these nuclei.

Finally, we assess the impact of using different nuclear models on the extraction of $V_{ud}$. We can extract $V_{ud}$ using the master formula of Eq.~(\ref{eq:master}). Adopting the values reported in Ref.~\cite{Hardy:2020qwl}, the value of $V_{ud}$ is,
\begin{equation}
 V_{ud} |^{\rm HT}_{^{10} \rm C} = 0.97318 (66)_{\rm exp} (9)_{\Delta^V_R} (24)_{\delta_{\rm NS}} (9)_{\rm \delta_C} \, ,
\end{equation}
Employing the values of $\delta_C$ computed in this work and the $\delta_{\rm NS}$ values of Ref.~\cite{King:2025akz}, we get the following values,

\begin{align}
 & \text{NV2+3-Ia}     &\left. V_{ud} \right|_{^{10} \rm C} &= 0.97338 (66)_{\rm exp} (9)_{\Delta^V_R}  (38)_{\delta_{\rm NS}} (77)_{\delta_C} \, ,\\
 &\text{AV18+UX} & \left. V_{ud} \right|_{^{10} \rm C} &=  0.97296 (66)_{\rm exp} (9)_{\Delta^V_R}  (23)_{\delta_{\rm NS}} (34)_{\delta_C} \, ,\\
 &\text{AV18+IL7} & \left. V_{ud} \right|_{^{10} \rm C} &=  0.97299 (66)_{\rm exp} (9)_{\Delta^V_R}  (31)_{\delta_{\rm NS}} (38)_{\delta_C} \, ,\\
 & \text{NV2+3-Ia$^{\star}$}     &\left. V_{ud} \right|_{^{10} \rm C} &= 0.97281 (66)_{\rm exp} (9)_{\Delta^V_R}  (32)_{\delta_{\rm NS}} (49)_{\delta_C} \, .
\end{align} 
If we use the values of $\delta_{\rm NS}$ and $\Delta^V_R$ from Ref.~\cite{Gennari:2024sbn} and $\delta_C$ from Ref.~\cite{Hardy:2020qwl}, we obtain,
\[
V_{ud}\big|_{^{10}\mathrm{C}}^{[12]}
=
0.97317(66)_{\mathrm{exp}}
(9)_{\Delta_R^V}
(16)_{\delta_{\rm NS}}
(9)_{\delta_C}.
\]
The range of values are compatible within the uncertainties. 

In the GFMC calculations, the uncertainty on $\delta_C$ provides the dominant theoretical error, followed by the uncertainty on $\delta_{\rm NS}$ that arises from the need to fit unknown coupling constants in the effective field theory approach of Refs.~\cite{Cirigliano:2024rfk,Cirigliano:2024msg}. In order to benefit CKM unitarity tests in the future, a more precise value of $\delta_C$ will be necessary. 
%\bob{Within the present QMC framework, VMC trial wave functions with larger initial $T=0$ components in the $^{10}$B$^*$ state might help, but correlations of the kind in Eq.(\ref{eq:ucib}) will tend to give equal amounts of $T=0$ and $T=2$, whereas the GFMC propagation clearly wants the former $\sim 3$ times larger.}
Using techniques to perform perturbative calculations in QMC developed in Ref.~\cite{Curry:2023mkm}, and the strategy laid out in Refs.~\cite{Seng:2022epj,Seng:2023cvt}, one could perform an alternate evaluation of $\delta_C$ that does not involve computing the small difference between two values. 

Finally, it is worth noting that while the overall theoretical uncertainty is increased by this determination of $\delta_C$, experiment remains the dominant source of uncertainty except in the case of the calculation using the NV2+3-Ia model. The experimental uncertainty is driven primarily by the measurement of the branching ratio for the superallowed decay~\cite{Hardy:2020qwl}. Improving upon the results in this work, as well as determining the unknowns that drive the uncertainty of $\delta_{\rm NS}$, would help to bring the overall error down. Quantum sensing techniques have been successful for performing precision spectroscopy of recoil ions after $\beta$ decay~\cite{BeEST:2020nns,Friedrich:2020nze}, and on-line measurements of recoil spectra at radioactive beam facilities using these approaches could help to reduce the uncertainty in the branching ratio and on $\mathcal{F}t$. Thus, such theoretical and experimental endeavors undertaken in tandem would improve the future determination of $\mathcal{F}t$ and sensitivity to new physics.

\section{Conclusion}
\label{sec:conclusions}

In this work, we present, to the best of our knowledge, the first {\it ab initio} QMC calculation of the isospin-symmetry-breaking correction $\delta_C$ for the superallowed $\beta$ decay of $^{10}$C. Using both phenomenological (AV18+UX, AV18+IL7) and chiral effective field theory (NV2+3-Ia and NV2+3-Ia$^\star$) Hamiltonians, we compute the Fermi matrix element within VMC and GFMC approaches and quantify its deviation from the canonical $\sqrt{2}$ value associated with exact isospin symmetry.

At the GFMC level, the Fermi matrix element deviates from $\sqrt{2}$ at the level of $10^{-3}$ for all Hamiltonians, corresponding to $\delta_C$ values in the range $\approx 0.15$--$0.25\%$. The individual results are statistically consistent, with all values overlapping within one standard deviation. Consequently, no statistically significant dependence on the choice of Hamiltonian can be established, although the spread in central values suggests a residual sensitivity to details of the nuclear interaction that is not yet quantitatively resolved.

The imaginary-time propagation of the Fermi matrix element shows variations across Hamiltonians. In all cases, the mixed estimates bracket $\sqrt{2}$, and the extrapolated GFMC results remain stable within the chosen averaging intervals. While differences in the $\tau$ dependence are observed, they do not provide a clear separation between phenomenological and chiral interactions.

The determination of $\delta_C$ is intrinsically challenging, as it relies on resolving small deviations of the Fermi matrix element from $\sqrt{2}$ and is therefore sensitive to cancellations in the mixed-estimator procedure. The resulting uncertainties vary across Hamiltonians, reflecting both the behavior of the mixed estimates and the extent of the imaginary-time intervals used for averaging. In particular, shorter propagation windows, such as those used for some of the NV2+3 interactions, limit the plateau region and lead to a less constrained extraction of $\delta_C$. This sensitivity is reflected in the relatively large fractional uncertainties, ranging from approximately $34\%$ to $65\%$, which limit the ability to draw more definitive conclusions regarding interaction dependence.

Within these uncertainties, the present results are consistent with existing evaluations of $\delta_C$ and yield values of $V_{ud}$ compatible with current determinations. However, within continuum QMC methods, the uncertainty in $\delta_C$ remains the dominant theoretical limitation for this system. In contrast, in the analyses of Hardy and Towner~\cite{Hardy:2014qxa,Hardy:2020qwl}, the uncertainty associated with $\delta_C$ is smaller than that of $\delta_{\rm NS}$.
Reducing it will require methodological developments, including alternative estimators or perturbative strategies that mitigate cancellations in the mixed-estimator procedure, as well as improvements in the trial wave function. In particular, incorporating the $T=0$ component that emerges during GFMC propagation more accurately at the VMC level may help stabilize the mixed estimates and enable a more precise determination of $\delta_C$.

Together with recent {\it ab initio} calculations of the radiative
correction $\delta_{\rm NS}$, this work represents a step toward a unified and systematically improvable description of nuclear-structure-dependent corrections in superallowed $\beta$ decay.  Despite current theoretical limitations, the uncertainty on $V_{ud}$ in $^{10}$C remains dominated by experiment, in particular by the measurement of the branching ratio. Future improvements in {\it ab initio} determinations of $\delta_C$, together with consistent evaluations of nuclear-structure-dependent radiative corrections, may help motivate more precise measurements of this quantity.

Coordinated progress in theory and experiment will be important for improving the determination of $\mathcal{F}t$ values and enhancing sensitivity to possible physics beyond the Standard Model. This is especially relevant for searches for non-standard scalar interactions, since the sensitivity to a possible Fierz interference term scales with the average inverse positron energy and is therefore enhanced in the lowest-$Z$ superallowed transitions.  As a result, $^{10}$C and $^{14}$O play an important role in setting bounds on scalar currents~\cite{Hardy:2020qwl,Dunlop:2016oif}, and improved control of the nuclear corrections in $^{10}$C can have a direct impact on present limits on such interactions.

\begin{acknowledgments}
We thank S.~C.~Pieper for early work on this problem. We are grateful to
Martín González and E.~Mereghetti for useful discussions during the
preparation of this manuscript. This work is supported by the U.S. Department of Energy, Office of Science, Office of Nuclear Physics, under the 2021 Early Career Award No. DE-SC0022002 (M.P.); under Contracts No. DE-SC0021027 (S.P.) and DE-AC02-06CH11357 (A.L. and R.B.W.). This work was also supported by the Office of Advanced Scientific Computing Research, Scientific Discovery through Advanced Computing (SciDAC) NUCLEI program (G.B.K., A.L., S.P., M.P., and R.B.W.) and by grant PID2023-147458NB-C21 (A.~L.) funded by MCIN/AEI/10.13039/501100011033, and by the European Union. Financial support from Los Alamos National Laboratory’s Laboratory Directed Research and Development program under project 20240742PRD1 (G.B.K.) is gratefully acknowledged. 
Los Alamos National Laboratory is operated by Triad National Security, LLC, for the National Nuclear Security Administration of U.S.\ Department of Energy (Contract No.\
89233218CNA000001). 
We acknowledge support from the DOE Topical Collaboration ``Nuclear Theory for New Physics,'' award No.\ DE-SC0023663.
This research used computational resources from the Laboratory Computing Resource Center of Argonne National Laboratory.
\end{acknowledgments}

\bibliography{delta}

\end{document}